%% file: amm821.tex
\documentclass[12pt]{article}
\begin{document}
\begin{center}

{\LARGE {\bf  The anomalous magnetic moment of muon: 
from the E821 experiment to bilepton masses}}\\[7mm]
{\bf Nguyen Anh Ky} and {\bf Hoang Ngoc Long}\\[2mm]
Institute of Physics, NCST\\
P.O. Box 429, Bo Ho, Hanoi 10000, Vietnam\\[1cm]
\end{center}

\begin{abstract}
   Bilepton masses in the 3-3-1 models and the recently announced 
 E821 experiment result are discussed. We get bounds on the bilepton masses 
243 GeV $\leq M_Y \leq$ 357  GeV in the minimal model and  
70 GeV $ \leq M_Y \leq$ 115 GeV in the model with right-handed neutrinos.
\\[4mm]
\end{abstract}

     The anomalous magnetic moments (AMM's) of electron and muon have been taken as one of the most 
precise and beautiful tests of the validity of quantum field theories like QED and the 
standard model (SM).  The accuracy of both the theoretical and experimental determination of these 
quantities is ceaselessly improved. Today they are calculated and measured to extreemly high 
precision. The surprising
agreement between the theoretical and experimental AMM of electron (AMME) 
$a_e\equiv (g_e-2)/2$ was a triumph of QED. Less precise than the AMME but about 
40,000 $\left( \approx m_\mu^2/m_e^2\right)$ times more sensitive to the quantum corrections 
and, therefore, more sensitive  to the ``New Physics" effects, the AMM of muon (AMMM) 
$a_\mu\equiv (g_\mu - 2)/2 $ is currently preferable in testing the SM. The recent E821 
experiment at BNL measuring the AMMM to a precision of 1.3 parts per million (pmm) gave 
a deviation from the SM theoretical value   
\begin{equation}
\delta a_\mu = 43  (16)\times 10^{-10},
\end{equation}
which is  2.6 times the normal deviation~\cite {brown}.
This result may be an indication of new physics and
has caused extensive interest in the current literature.
In this Note, only contributions from the 
 models based on the  $\mbox{SU}(3)_C \times \mbox{SU}(3)_L\times
\mbox{U}(1)$ ( 3-3-1) gauge group~\cite{m331,r331}, or more precisely, 
only bilepton contributions of  these models, 
to the AMMM are considered and compared with the recently announced result of the 
E821 experiment. The constributions from the bileptons  should be much bigger than those of 
the scalar sector which could be neglected for the moment\footnote{A special case for a 
contribution from the scalar sector in the minimal model is considered in ~\cite{pires}.}. 

   In a recent paper \cite{kls} the AMMM's  were calulated for both versions of 
the  3- 3- 1 models, the minimal version (MV)~\cite{m331} and 
the model with right-handed neutrinos (MRHN),  
but analysed in the light of previous, less precise, experimental data.  At the present 
value $a_\mu^{exp}=11~659~202 ~(14)(6)\times 10^{-10}$ of the experimental AMMM,
 the minimal version of the 3-3-1 model could be a dominating  ``New Physics", while the 
MRHN, if allowed by the Nature, gives a very small contribution to the AMMM.
Analysing the bilepton masses  and the new experimental data  we guess that 
 $a_\mu^{exp}$ would become closer to the theoretical AMMM ~$a_\mu^{SM}=
11~659~159.6 ~(6.7)\times 10^{-10}$ (0.57 ppm) in the future measurements.  
Then the MRHN may play a more significant role in generating the AMM's and 
phenomena beyond the SM. 

     Contributions to the AMMM from new gauge bosons in the MV\\
\begin{equation}
\delta a_\mu^{tm} = \frac{G_F m_W^2 m_\mu^2}{\sqrt{2}\pi^2}
\left[ \frac{23}{4 M_Y^2} +
 \frac{2 (1-4 s_W^2)}{9 c_W^2
M_{Z'}^2} \right]
\label{dgzmtt}
\end{equation}
\\
and in the MRHN \\
\begin{equation}
\delta a_\mu^{tr} = \frac{G_F m_W^2 m_\mu^2}{12\sqrt{2}\pi^2}
\left\{ \frac{5}{M_Y^2} -  \frac{[5-(1-4 s_W^2)^2]}{2 c_W^2
(3 - 4 s_W^2)M_{Z'}^2} \right\}
\label{dgzrt}
\end{equation}
\\
are given in~\cite{kls} where  $s_W\equiv \sin\theta_W,  c_W\equiv \cos\theta_W$ 
and $m_\mu$,  $M_Y$ and $M_{Z'}$ are 
masses of the muon, the bileptons $Y$ and the $Z'$, respectively. 
Theses contributions  $\delta a_\mu^{tm} $ and $\delta a_\mu^{tr} $ are plotted in Fig. 1 
and Fig. 2  as functions of $M_Y$ for given $M_{Z'}$. 
The horizontal lines in the figures correspond to the upper and lower limits on the 
right-handed side of Eq. (1). From the data on the atomic parity violation in cesium it follows that  the $Z'$ mass  may take a value of 2000 GeV in the MV and 2200 GeV in the MRHN~\cite{lt}.  As seen from the figures, the mass bounds 243 GeV  $\leq M_Y \leq$ 357 GeV ensuring the MV to contribute to the AMMM a value comparable to the present $\delta a_\mu$  are quite reasonable (see Fig.1), while the mass bounds  70 GeV $\leq M_Y \leq$ 115 GeV ensuring the MRHN to give such a contribution are too small (see Fig. 2). We note that  the mass bounds derived here for the MV do not contradict the condition $M_Y \geq 230$ GeV followed from the wrong muon decay.  The mass bounds for the MRHN, however, are still under the lower limit 230 GeV. Moreover, on this mass limit
the contribution of the MRHN to the AMMM cannot exceed $13.1\times 10^{-10}$ 
which is considerably smaller than the present experimetal deviation  
$\delta a_\mu = 43 (16)\times 10^{-10}$ from the theoretical  value $a_\mu^{SM}$. 
So, according  to the latest experimental data the MV seems to be  a more  
suitable model of the physics beyond the SM than the MRHN. 

   To finish this Note we conclude that new gauge bosons in the MV can provide a contribution
consistent with the latest experimental result on AMMM which, however, disfavours  the MRHN. 
For the latter  the bounds obtained here are lower than that followed from the wrong muon 
decay. With  $M_Y$ around 230 GeV, the new gauge bosons in the MRHN provide a contribution of just  
one third of the present experimental result. 

{\bf Acknowledgements}: This work was supported in part 
by the National Research Program 
for Natural Sciences of Vietnam under Grant No.  KT-04.1.2

\newpage 
\begin{center}
\input {fig821-1.tex}
\\[5mm]
Fig. 1:  $\delta a_\mu^{tm}$ as a function of $M_{Y}$
with  $M_{Z'}= 2000 $ GeV.
\end{center}
\vspace*{1cm}
\begin{center}
\input {fig821-2.tex}
\\[5mm]
Fig. 2: $\delta a_\mu^{tr}$ as a function of $M_{Y}$
with $M_{Z'}= 2200 $ GeV. 
\end{center}
\end{document}

%% file: fig821-1.tex
\setlength{\unitlength}{0.240900pt}
\ifx\plotpoint\undefined\newsavebox{\plotpoint}\fi
\sbox{\plotpoint}{\rule[-0.200pt]{0.400pt}{0.400pt}}%
\begin{picture}(1500,900)(0,0)
\font\gnuplot=cmr10 at 10pt
\gnuplot
\sbox{\plotpoint}{\rule[-0.200pt]{1.400pt}{1.400pt}}%
\put(220.0,198.0){\rule[-0.200pt]{292.934pt}{0.400pt}}
\put(220.0,198.0){\rule[-0.200pt]{4.818pt}{0.400pt}}
\put(198,198){\makebox(0,0)[r]{0}}
\put(1416.0,198.0){\rule[-0.200pt]{4.818pt}{0.400pt}}
\put(220.0,368.0){\rule[-0.200pt]{4.818pt}{0.400pt}}
\put(198,368){\makebox(0,0)[r]{0.5}}
\put(1416.0,368.0){\rule[-0.200pt]{4.818pt}{0.400pt}}
\put(220.0,537.0){\rule[-0.200pt]{4.818pt}{0.400pt}}
\put(198,537){\makebox(0,0)[r]{1}}
\put(1416.0,537.0){\rule[-0.200pt]{4.818pt}{0.400pt}}
\put(220.0,707.0){\rule[-0.200pt]{4.818pt}{0.400pt}}
\put(198,707){\makebox(0,0)[r]{1.5}}
\put(1416.0,707.0){\rule[-0.200pt]{4.818pt}{0.400pt}}
\put(220.0,877.0){\rule[-0.200pt]{4.818pt}{0.400pt}}
\put(198,877){\makebox(0,0)[r]{2}}
\put(1416.0,877.0){\rule[-0.200pt]{4.818pt}{0.400pt}}
\put(220.0,113.0){\rule[-0.200pt]{0.400pt}{4.818pt}}
\put(220,68){\makebox(0,0){100}}
\put(220.0,857.0){\rule[-0.200pt]{0.400pt}{4.818pt}}
\put(342.0,113.0){\rule[-0.200pt]{0.400pt}{4.818pt}}
\put(342,68){\makebox(0,0){150}}
\put(342.0,857.0){\rule[-0.200pt]{0.400pt}{4.818pt}}
\put(463.0,113.0){\rule[-0.200pt]{0.400pt}{4.818pt}}
\put(463,68){\makebox(0,0){200}}
\put(463.0,857.0){\rule[-0.200pt]{0.400pt}{4.818pt}}
\put(585.0,113.0){\rule[-0.200pt]{0.400pt}{4.818pt}}
\put(585,68){\makebox(0,0){250}}
\put(585.0,857.0){\rule[-0.200pt]{0.400pt}{4.818pt}}
\put(706.0,113.0){\rule[-0.200pt]{0.400pt}{4.818pt}}
\put(706,68){\makebox(0,0){300}}
\put(706.0,857.0){\rule[-0.200pt]{0.400pt}{4.818pt}}
\put(828.0,113.0){\rule[-0.200pt]{0.400pt}{4.818pt}}
\put(828,68){\makebox(0,0){350}}
\put(828.0,857.0){\rule[-0.200pt]{0.400pt}{4.818pt}}
\put(950.0,113.0){\rule[-0.200pt]{0.400pt}{4.818pt}}
\put(950,68){\makebox(0,0){400}}
\put(950.0,857.0){\rule[-0.200pt]{0.400pt}{4.818pt}}
\put(1071.0,113.0){\rule[-0.200pt]{0.400pt}{4.818pt}}
\put(1071,68){\makebox(0,0){450}}
\put(1071.0,857.0){\rule[-0.200pt]{0.400pt}{4.818pt}}
\put(1193.0,113.0){\rule[-0.200pt]{0.400pt}{4.818pt}}
\put(1193,68){\makebox(0,0){500}}
\put(1193.0,857.0){\rule[-0.200pt]{0.400pt}{4.818pt}}
\put(1314.0,113.0){\rule[-0.200pt]{0.400pt}{4.818pt}}
\put(1314,68){\makebox(0,0){550}}
\put(1314.0,857.0){\rule[-0.200pt]{0.400pt}{4.818pt}}
\put(1436.0,113.0){\rule[-0.200pt]{0.400pt}{4.818pt}}
\put(1436,68){\makebox(0,0){600}}
\put(1436.0,857.0){\rule[-0.200pt]{0.400pt}{4.818pt}}
\put(220.0,113.0){\rule[-0.200pt]{292.934pt}{0.400pt}}
\put(1436.0,113.0){\rule[-0.200pt]{0.400pt}{184.048pt}}
\put(220.0,877.0){\rule[-0.200pt]{292.934pt}{0.400pt}}
\put(20,495){\makebox(0,0){{$\delta a_{\mu}^{tm} \times 10^{-8}$}}}
\put(828,-20){\makebox(0,0){$M_Y$(GeV)}}
\put(220.0,113.0){\rule[-0.200pt]{0.400pt}{184.048pt}}
\put(1306,812){\makebox(0,0)[r]{{$\delta a_{\mu}^{tm}$}}}
\multiput(297.59,869.94)(0.489,-2.067){15}{\rule{0.118pt}{1.700pt}}
\multiput(296.17,873.47)(9.000,-32.472){2}{\rule{0.400pt}{0.850pt}}
\multiput(306.58,834.36)(0.492,-1.918){21}{\rule{0.119pt}{1.600pt}}
\multiput(305.17,837.68)(12.000,-41.679){2}{\rule{0.400pt}{0.800pt}}
\multiput(318.58,790.35)(0.493,-1.607){23}{\rule{0.119pt}{1.362pt}}
\multiput(317.17,793.17)(13.000,-38.174){2}{\rule{0.400pt}{0.681pt}}
\multiput(331.58,749.47)(0.492,-1.573){21}{\rule{0.119pt}{1.333pt}}
\multiput(330.17,752.23)(12.000,-34.233){2}{\rule{0.400pt}{0.667pt}}
\multiput(343.58,713.02)(0.492,-1.401){21}{\rule{0.119pt}{1.200pt}}
\multiput(342.17,715.51)(12.000,-30.509){2}{\rule{0.400pt}{0.600pt}}
\multiput(355.58,680.43)(0.492,-1.272){21}{\rule{0.119pt}{1.100pt}}
\multiput(354.17,682.72)(12.000,-27.717){2}{\rule{0.400pt}{0.550pt}}
\multiput(367.58,651.01)(0.493,-1.091){23}{\rule{0.119pt}{0.962pt}}
\multiput(366.17,653.00)(13.000,-26.004){2}{\rule{0.400pt}{0.481pt}}
\multiput(380.58,623.13)(0.492,-1.056){21}{\rule{0.119pt}{0.933pt}}
\multiput(379.17,625.06)(12.000,-23.063){2}{\rule{0.400pt}{0.467pt}}
\multiput(392.58,598.40)(0.492,-0.970){21}{\rule{0.119pt}{0.867pt}}
\multiput(391.17,600.20)(12.000,-21.201){2}{\rule{0.400pt}{0.433pt}}
\multiput(404.58,575.90)(0.493,-0.814){23}{\rule{0.119pt}{0.746pt}}
\multiput(403.17,577.45)(13.000,-19.451){2}{\rule{0.400pt}{0.373pt}}
\multiput(417.58,554.96)(0.492,-0.798){21}{\rule{0.119pt}{0.733pt}}
\multiput(416.17,556.48)(12.000,-17.478){2}{\rule{0.400pt}{0.367pt}}
\multiput(429.58,536.09)(0.492,-0.755){21}{\rule{0.119pt}{0.700pt}}
\multiput(428.17,537.55)(12.000,-16.547){2}{\rule{0.400pt}{0.350pt}}
\multiput(441.58,518.37)(0.492,-0.669){21}{\rule{0.119pt}{0.633pt}}
\multiput(440.17,519.69)(12.000,-14.685){2}{\rule{0.400pt}{0.317pt}}
\multiput(453.58,502.67)(0.493,-0.576){23}{\rule{0.119pt}{0.562pt}}
\multiput(452.17,503.83)(13.000,-13.834){2}{\rule{0.400pt}{0.281pt}}
\multiput(466.58,487.51)(0.492,-0.625){21}{\rule{0.119pt}{0.600pt}}
\multiput(465.17,488.75)(12.000,-13.755){2}{\rule{0.400pt}{0.300pt}}
\multiput(478.58,472.79)(0.492,-0.539){21}{\rule{0.119pt}{0.533pt}}
\multiput(477.17,473.89)(12.000,-11.893){2}{\rule{0.400pt}{0.267pt}}
\multiput(490.00,460.92)(0.539,-0.492){21}{\rule{0.533pt}{0.119pt}}
\multiput(490.00,461.17)(11.893,-12.000){2}{\rule{0.267pt}{0.400pt}}
\multiput(503.00,448.92)(0.543,-0.492){19}{\rule{0.536pt}{0.118pt}}
\multiput(503.00,449.17)(10.887,-11.000){2}{\rule{0.268pt}{0.400pt}}
\multiput(515.00,437.92)(0.543,-0.492){19}{\rule{0.536pt}{0.118pt}}
\multiput(515.00,438.17)(10.887,-11.000){2}{\rule{0.268pt}{0.400pt}}
\multiput(527.00,426.92)(0.600,-0.491){17}{\rule{0.580pt}{0.118pt}}
\multiput(527.00,427.17)(10.796,-10.000){2}{\rule{0.290pt}{0.400pt}}
\multiput(539.00,416.93)(0.728,-0.489){15}{\rule{0.678pt}{0.118pt}}
\multiput(539.00,417.17)(11.593,-9.000){2}{\rule{0.339pt}{0.400pt}}
\multiput(552.00,407.93)(0.669,-0.489){15}{\rule{0.633pt}{0.118pt}}
\multiput(552.00,408.17)(10.685,-9.000){2}{\rule{0.317pt}{0.400pt}}
\multiput(564.00,398.93)(0.758,-0.488){13}{\rule{0.700pt}{0.117pt}}
\multiput(564.00,399.17)(10.547,-8.000){2}{\rule{0.350pt}{0.400pt}}
\multiput(576.00,390.93)(0.758,-0.488){13}{\rule{0.700pt}{0.117pt}}
\multiput(576.00,391.17)(10.547,-8.000){2}{\rule{0.350pt}{0.400pt}}
\multiput(588.00,382.93)(0.950,-0.485){11}{\rule{0.843pt}{0.117pt}}
\multiput(588.00,383.17)(11.251,-7.000){2}{\rule{0.421pt}{0.400pt}}
\multiput(601.00,375.93)(0.874,-0.485){11}{\rule{0.786pt}{0.117pt}}
\multiput(601.00,376.17)(10.369,-7.000){2}{\rule{0.393pt}{0.400pt}}
\multiput(613.00,368.93)(1.033,-0.482){9}{\rule{0.900pt}{0.116pt}}
\multiput(613.00,369.17)(10.132,-6.000){2}{\rule{0.450pt}{0.400pt}}
\multiput(625.00,362.93)(1.123,-0.482){9}{\rule{0.967pt}{0.116pt}}
\multiput(625.00,363.17)(10.994,-6.000){2}{\rule{0.483pt}{0.400pt}}
\multiput(638.00,356.93)(1.033,-0.482){9}{\rule{0.900pt}{0.116pt}}
\multiput(638.00,357.17)(10.132,-6.000){2}{\rule{0.450pt}{0.400pt}}
\multiput(650.00,350.93)(1.033,-0.482){9}{\rule{0.900pt}{0.116pt}}
\multiput(650.00,351.17)(10.132,-6.000){2}{\rule{0.450pt}{0.400pt}}
\multiput(662.00,344.93)(1.267,-0.477){7}{\rule{1.060pt}{0.115pt}}
\multiput(662.00,345.17)(9.800,-5.000){2}{\rule{0.530pt}{0.400pt}}
\multiput(674.00,339.93)(1.378,-0.477){7}{\rule{1.140pt}{0.115pt}}
\multiput(674.00,340.17)(10.634,-5.000){2}{\rule{0.570pt}{0.400pt}}
\multiput(687.00,334.94)(1.651,-0.468){5}{\rule{1.300pt}{0.113pt}}
\multiput(687.00,335.17)(9.302,-4.000){2}{\rule{0.650pt}{0.400pt}}
\multiput(699.00,330.93)(1.267,-0.477){7}{\rule{1.060pt}{0.115pt}}
\multiput(699.00,331.17)(9.800,-5.000){2}{\rule{0.530pt}{0.400pt}}
\multiput(711.00,325.94)(1.797,-0.468){5}{\rule{1.400pt}{0.113pt}}
\multiput(711.00,326.17)(10.094,-4.000){2}{\rule{0.700pt}{0.400pt}}
\multiput(724.00,321.94)(1.651,-0.468){5}{\rule{1.300pt}{0.113pt}}
\multiput(724.00,322.17)(9.302,-4.000){2}{\rule{0.650pt}{0.400pt}}
\multiput(736.00,317.94)(1.651,-0.468){5}{\rule{1.300pt}{0.113pt}}
\multiput(736.00,318.17)(9.302,-4.000){2}{\rule{0.650pt}{0.400pt}}
\multiput(748.00,313.94)(1.651,-0.468){5}{\rule{1.300pt}{0.113pt}}
\multiput(748.00,314.17)(9.302,-4.000){2}{\rule{0.650pt}{0.400pt}}
\multiput(760.00,309.95)(2.695,-0.447){3}{\rule{1.833pt}{0.108pt}}
\multiput(760.00,310.17)(9.195,-3.000){2}{\rule{0.917pt}{0.400pt}}
\multiput(773.00,306.95)(2.472,-0.447){3}{\rule{1.700pt}{0.108pt}}
\multiput(773.00,307.17)(8.472,-3.000){2}{\rule{0.850pt}{0.400pt}}
\multiput(785.00,303.94)(1.651,-0.468){5}{\rule{1.300pt}{0.113pt}}
\multiput(785.00,304.17)(9.302,-4.000){2}{\rule{0.650pt}{0.400pt}}
\multiput(797.00,299.95)(2.695,-0.447){3}{\rule{1.833pt}{0.108pt}}
\multiput(797.00,300.17)(9.195,-3.000){2}{\rule{0.917pt}{0.400pt}}
\put(810,296.17){\rule{2.500pt}{0.400pt}}
\multiput(810.00,297.17)(6.811,-2.000){2}{\rule{1.250pt}{0.400pt}}
\multiput(822.00,294.95)(2.472,-0.447){3}{\rule{1.700pt}{0.108pt}}
\multiput(822.00,295.17)(8.472,-3.000){2}{\rule{0.850pt}{0.400pt}}
\multiput(834.00,291.95)(2.472,-0.447){3}{\rule{1.700pt}{0.108pt}}
\multiput(834.00,292.17)(8.472,-3.000){2}{\rule{0.850pt}{0.400pt}}
\put(846,288.17){\rule{2.700pt}{0.400pt}}
\multiput(846.00,289.17)(7.396,-2.000){2}{\rule{1.350pt}{0.400pt}}
\multiput(859.00,286.95)(2.472,-0.447){3}{\rule{1.700pt}{0.108pt}}
\multiput(859.00,287.17)(8.472,-3.000){2}{\rule{0.850pt}{0.400pt}}
\put(871,283.17){\rule{2.500pt}{0.400pt}}
\multiput(871.00,284.17)(6.811,-2.000){2}{\rule{1.250pt}{0.400pt}}
\multiput(883.00,281.95)(2.695,-0.447){3}{\rule{1.833pt}{0.108pt}}
\multiput(883.00,282.17)(9.195,-3.000){2}{\rule{0.917pt}{0.400pt}}
\put(896,278.17){\rule{2.500pt}{0.400pt}}
\multiput(896.00,279.17)(6.811,-2.000){2}{\rule{1.250pt}{0.400pt}}
\put(908,276.17){\rule{2.500pt}{0.400pt}}
\multiput(908.00,277.17)(6.811,-2.000){2}{\rule{1.250pt}{0.400pt}}
\put(920,274.17){\rule{2.500pt}{0.400pt}}
\multiput(920.00,275.17)(6.811,-2.000){2}{\rule{1.250pt}{0.400pt}}
\put(932,272.17){\rule{2.700pt}{0.400pt}}
\multiput(932.00,273.17)(7.396,-2.000){2}{\rule{1.350pt}{0.400pt}}
\put(945,270.17){\rule{2.500pt}{0.400pt}}
\multiput(945.00,271.17)(6.811,-2.000){2}{\rule{1.250pt}{0.400pt}}
\put(957,268.67){\rule{2.891pt}{0.400pt}}
\multiput(957.00,269.17)(6.000,-1.000){2}{\rule{1.445pt}{0.400pt}}
\put(969,267.17){\rule{2.700pt}{0.400pt}}
\multiput(969.00,268.17)(7.396,-2.000){2}{\rule{1.350pt}{0.400pt}}
\put(982,265.17){\rule{2.500pt}{0.400pt}}
\multiput(982.00,266.17)(6.811,-2.000){2}{\rule{1.250pt}{0.400pt}}
\put(994,263.67){\rule{2.891pt}{0.400pt}}
\multiput(994.00,264.17)(6.000,-1.000){2}{\rule{1.445pt}{0.400pt}}
\put(1006,262.17){\rule{2.500pt}{0.400pt}}
\multiput(1006.00,263.17)(6.811,-2.000){2}{\rule{1.250pt}{0.400pt}}
\put(1018,260.67){\rule{3.132pt}{0.400pt}}
\multiput(1018.00,261.17)(6.500,-1.000){2}{\rule{1.566pt}{0.400pt}}
\put(1031,259.17){\rule{2.500pt}{0.400pt}}
\multiput(1031.00,260.17)(6.811,-2.000){2}{\rule{1.250pt}{0.400pt}}
\put(1043,257.67){\rule{2.891pt}{0.400pt}}
\multiput(1043.00,258.17)(6.000,-1.000){2}{\rule{1.445pt}{0.400pt}}
\put(1055,256.17){\rule{2.700pt}{0.400pt}}
\multiput(1055.00,257.17)(7.396,-2.000){2}{\rule{1.350pt}{0.400pt}}
\put(1068,254.67){\rule{2.891pt}{0.400pt}}
\multiput(1068.00,255.17)(6.000,-1.000){2}{\rule{1.445pt}{0.400pt}}
\put(1080,253.67){\rule{2.891pt}{0.400pt}}
\multiput(1080.00,254.17)(6.000,-1.000){2}{\rule{1.445pt}{0.400pt}}
\put(1092,252.67){\rule{2.891pt}{0.400pt}}
\multiput(1092.00,253.17)(6.000,-1.000){2}{\rule{1.445pt}{0.400pt}}
\put(1104,251.67){\rule{3.132pt}{0.400pt}}
\multiput(1104.00,252.17)(6.500,-1.000){2}{\rule{1.566pt}{0.400pt}}
\put(1117,250.17){\rule{2.500pt}{0.400pt}}
\multiput(1117.00,251.17)(6.811,-2.000){2}{\rule{1.250pt}{0.400pt}}
\put(1129,248.67){\rule{2.891pt}{0.400pt}}
\multiput(1129.00,249.17)(6.000,-1.000){2}{\rule{1.445pt}{0.400pt}}
\put(1141,247.67){\rule{2.891pt}{0.400pt}}
\multiput(1141.00,248.17)(6.000,-1.000){2}{\rule{1.445pt}{0.400pt}}
\put(1153,246.67){\rule{3.132pt}{0.400pt}}
\multiput(1153.00,247.17)(6.500,-1.000){2}{\rule{1.566pt}{0.400pt}}
\put(1166,245.67){\rule{2.891pt}{0.400pt}}
\multiput(1166.00,246.17)(6.000,-1.000){2}{\rule{1.445pt}{0.400pt}}
\put(1178,244.67){\rule{2.891pt}{0.400pt}}
\multiput(1178.00,245.17)(6.000,-1.000){2}{\rule{1.445pt}{0.400pt}}
\put(1190,243.67){\rule{3.132pt}{0.400pt}}
\multiput(1190.00,244.17)(6.500,-1.000){2}{\rule{1.566pt}{0.400pt}}
\put(1203,242.67){\rule{2.891pt}{0.400pt}}
\multiput(1203.00,243.17)(6.000,-1.000){2}{\rule{1.445pt}{0.400pt}}
\put(1215,241.67){\rule{2.891pt}{0.400pt}}
\multiput(1215.00,242.17)(6.000,-1.000){2}{\rule{1.445pt}{0.400pt}}
\put(1328.0,812.0){\rule[-0.200pt]{15.899pt}{0.400pt}}
\put(1239,240.67){\rule{3.132pt}{0.400pt}}
\multiput(1239.00,241.17)(6.500,-1.000){2}{\rule{1.566pt}{0.400pt}}
\put(1252,239.67){\rule{2.891pt}{0.400pt}}
\multiput(1252.00,240.17)(6.000,-1.000){2}{\rule{1.445pt}{0.400pt}}
\put(1264,238.67){\rule{2.891pt}{0.400pt}}
\multiput(1264.00,239.17)(6.000,-1.000){2}{\rule{1.445pt}{0.400pt}}
\put(1276,237.67){\rule{3.132pt}{0.400pt}}
\multiput(1276.00,238.17)(6.500,-1.000){2}{\rule{1.566pt}{0.400pt}}
\put(1227.0,242.0){\rule[-0.200pt]{2.891pt}{0.400pt}}
\put(1301,236.67){\rule{2.891pt}{0.400pt}}
\multiput(1301.00,237.17)(6.000,-1.000){2}{\rule{1.445pt}{0.400pt}}
\put(1313,235.67){\rule{2.891pt}{0.400pt}}
\multiput(1313.00,236.17)(6.000,-1.000){2}{\rule{1.445pt}{0.400pt}}
\put(1289.0,238.0){\rule[-0.200pt]{2.891pt}{0.400pt}}
\put(1338,234.67){\rule{2.891pt}{0.400pt}}
\multiput(1338.00,235.17)(6.000,-1.000){2}{\rule{1.445pt}{0.400pt}}
\put(1350,233.67){\rule{2.891pt}{0.400pt}}
\multiput(1350.00,234.17)(6.000,-1.000){2}{\rule{1.445pt}{0.400pt}}
\put(1325.0,236.0){\rule[-0.200pt]{3.132pt}{0.400pt}}
\put(1375,232.67){\rule{2.891pt}{0.400pt}}
\multiput(1375.00,233.17)(6.000,-1.000){2}{\rule{1.445pt}{0.400pt}}
\put(1387,231.67){\rule{2.891pt}{0.400pt}}
\multiput(1387.00,232.17)(6.000,-1.000){2}{\rule{1.445pt}{0.400pt}}
\put(1362.0,234.0){\rule[-0.200pt]{3.132pt}{0.400pt}}
\put(1411,230.67){\rule{3.132pt}{0.400pt}}
\multiput(1411.00,231.17)(6.500,-1.000){2}{\rule{1.566pt}{0.400pt}}
\put(1399.0,232.0){\rule[-0.200pt]{2.891pt}{0.400pt}}
\put(1424.0,231.0){\rule[-0.200pt]{2.891pt}{0.400pt}}

\put(1306,767){\makebox(0,0)[r]{Upper limit}}
\multiput(1328,767)(20.756,0.000){4}{\usebox{\plotpoint}}
\put(1394,767){\usebox{\plotpoint}}
\put(220,398){\usebox{\plotpoint}}
\put(220.00,398.00){\usebox{\plotpoint}}
\put(240.76,398.00){\usebox{\plotpoint}}
\multiput(245,398)(20.756,0.000){0}{\usebox{\plotpoint}}
\put(261.51,398.00){\usebox{\plotpoint}}
\multiput(269,398)(20.756,0.000){0}{\usebox{\plotpoint}}
\put(282.27,398.00){\usebox{\plotpoint}}
\put(303.02,398.00){\usebox{\plotpoint}}
\multiput(306,398)(20.756,0.000){0}{\usebox{\plotpoint}}
\put(323.78,398.00){\usebox{\plotpoint}}
\multiput(331,398)(20.756,0.000){0}{\usebox{\plotpoint}}
\put(344.53,398.00){\usebox{\plotpoint}}
\put(365.29,398.00){\usebox{\plotpoint}}
\multiput(367,398)(20.756,0.000){0}{\usebox{\plotpoint}}
\put(386.04,398.00){\usebox{\plotpoint}}
\multiput(392,398)(20.756,0.000){0}{\usebox{\plotpoint}}
\put(406.80,398.00){\usebox{\plotpoint}}
\put(427.55,398.00){\usebox{\plotpoint}}
\multiput(429,398)(20.756,0.000){0}{\usebox{\plotpoint}}
\put(448.31,398.00){\usebox{\plotpoint}}
\multiput(453,398)(20.756,0.000){0}{\usebox{\plotpoint}}
\put(469.07,398.00){\usebox{\plotpoint}}
\put(489.82,398.00){\usebox{\plotpoint}}
\multiput(490,398)(20.756,0.000){0}{\usebox{\plotpoint}}
\put(510.58,398.00){\usebox{\plotpoint}}
\multiput(515,398)(20.756,0.000){0}{\usebox{\plotpoint}}
\put(531.33,398.00){\usebox{\plotpoint}}
\multiput(539,398)(20.756,0.000){0}{\usebox{\plotpoint}}
\put(552.09,398.00){\usebox{\plotpoint}}
\put(572.84,398.00){\usebox{\plotpoint}}
\multiput(576,398)(20.756,0.000){0}{\usebox{\plotpoint}}
\put(593.60,398.00){\usebox{\plotpoint}}
\multiput(601,398)(20.756,0.000){0}{\usebox{\plotpoint}}
\put(614.35,398.00){\usebox{\plotpoint}}
\put(635.11,398.00){\usebox{\plotpoint}}
\multiput(638,398)(20.756,0.000){0}{\usebox{\plotpoint}}
\put(655.87,398.00){\usebox{\plotpoint}}
\multiput(662,398)(20.756,0.000){0}{\usebox{\plotpoint}}
\put(676.62,398.00){\usebox{\plotpoint}}
\put(697.38,398.00){\usebox{\plotpoint}}
\multiput(699,398)(20.756,0.000){0}{\usebox{\plotpoint}}
\put(718.13,398.00){\usebox{\plotpoint}}
\multiput(724,398)(20.756,0.000){0}{\usebox{\plotpoint}}
\put(738.89,398.00){\usebox{\plotpoint}}
\put(759.64,398.00){\usebox{\plotpoint}}
\multiput(760,398)(20.756,0.000){0}{\usebox{\plotpoint}}
\put(780.40,398.00){\usebox{\plotpoint}}
\multiput(785,398)(20.756,0.000){0}{\usebox{\plotpoint}}
\put(801.15,398.00){\usebox{\plotpoint}}
\put(821.91,398.00){\usebox{\plotpoint}}
\multiput(822,398)(20.756,0.000){0}{\usebox{\plotpoint}}
\put(842.66,398.00){\usebox{\plotpoint}}
\multiput(846,398)(20.756,0.000){0}{\usebox{\plotpoint}}
\put(863.42,398.00){\usebox{\plotpoint}}
\multiput(871,398)(20.756,0.000){0}{\usebox{\plotpoint}}
\put(884.18,398.00){\usebox{\plotpoint}}
\put(904.93,398.00){\usebox{\plotpoint}}
\multiput(908,398)(20.756,0.000){0}{\usebox{\plotpoint}}
\put(925.69,398.00){\usebox{\plotpoint}}
\multiput(932,398)(20.756,0.000){0}{\usebox{\plotpoint}}
\put(946.44,398.00){\usebox{\plotpoint}}
\put(967.20,398.00){\usebox{\plotpoint}}
\multiput(969,398)(20.756,0.000){0}{\usebox{\plotpoint}}
\put(987.95,398.00){\usebox{\plotpoint}}
\multiput(994,398)(20.756,0.000){0}{\usebox{\plotpoint}}
\put(1008.71,398.00){\usebox{\plotpoint}}
\put(1029.46,398.00){\usebox{\plotpoint}}
\multiput(1031,398)(20.756,0.000){0}{\usebox{\plotpoint}}
\put(1050.22,398.00){\usebox{\plotpoint}}
\multiput(1055,398)(20.756,0.000){0}{\usebox{\plotpoint}}
\put(1070.98,398.00){\usebox{\plotpoint}}
\put(1091.73,398.00){\usebox{\plotpoint}}
\multiput(1092,398)(20.756,0.000){0}{\usebox{\plotpoint}}
\put(1112.49,398.00){\usebox{\plotpoint}}
\multiput(1117,398)(20.756,0.000){0}{\usebox{\plotpoint}}
\put(1133.24,398.00){\usebox{\plotpoint}}
\multiput(1141,398)(20.756,0.000){0}{\usebox{\plotpoint}}
\put(1154.00,398.00){\usebox{\plotpoint}}
\put(1174.75,398.00){\usebox{\plotpoint}}
\multiput(1178,398)(20.756,0.000){0}{\usebox{\plotpoint}}
\put(1195.51,398.00){\usebox{\plotpoint}}
\multiput(1203,398)(20.756,0.000){0}{\usebox{\plotpoint}}
\put(1216.26,398.00){\usebox{\plotpoint}}
\put(1237.02,398.00){\usebox{\plotpoint}}
\multiput(1239,398)(20.756,0.000){0}{\usebox{\plotpoint}}
\put(1257.77,398.00){\usebox{\plotpoint}}
\multiput(1264,398)(20.756,0.000){0}{\usebox{\plotpoint}}
\put(1278.53,398.00){\usebox{\plotpoint}}
\put(1299.29,398.00){\usebox{\plotpoint}}
\multiput(1301,398)(20.756,0.000){0}{\usebox{\plotpoint}}
\put(1320.04,398.00){\usebox{\plotpoint}}
\multiput(1325,398)(20.756,0.000){0}{\usebox{\plotpoint}}
\put(1340.80,398.00){\usebox{\plotpoint}}
\put(1361.55,398.00){\usebox{\plotpoint}}
\multiput(1362,398)(20.756,0.000){0}{\usebox{\plotpoint}}
\put(1382.31,398.00){\usebox{\plotpoint}}
\multiput(1387,398)(20.756,0.000){0}{\usebox{\plotpoint}}
\put(1403.06,398.00){\usebox{\plotpoint}}
\put(1423.82,398.00){\usebox{\plotpoint}}
\multiput(1424,398)(20.756,0.000){0}{\usebox{\plotpoint}}
\put(1436,398){\usebox{\plotpoint}}
\sbox{\plotpoint}{\rule[-0.400pt]{0.800pt}{0.800pt}}%
\put(1306,722){\makebox(0,0)[r]{Lower limit}}
\put(1328.0,722.0){\rule[-0.400pt]{15.899pt}{0.800pt}}
\put(220,290){\usebox{\plotpoint}}
\put(220.0,290.0){\rule[-0.400pt]{292.934pt}{0.800pt}}
\end{picture}

%% file: fig821-2.tex
\setlength{\unitlength}{0.240900pt}
\ifx\plotpoint\undefined\newsavebox{\plotpoint}\fi
\sbox{\plotpoint}{\rule[-0.200pt]{0.400pt}{0.400pt}}%
\begin{picture}(1500,900)(0,0)
\font\gnuplot=cmr10 at 10pt
\gnuplot
\sbox{\plotpoint}{\rule[-0.200pt]{1.400pt}{1.400pt}}%
\put(220.0,113.0){\rule[-0.200pt]{292.934pt}{0.400pt}}
\put(220.0,113.0){\rule[-0.200pt]{4.818pt}{0.400pt}}
\put(198,113){\makebox(0,0)[r]{0}}
\put(1416.0,113.0){\rule[-0.200pt]{4.818pt}{0.400pt}}
\put(220.0,266.0){\rule[-0.200pt]{4.818pt}{0.400pt}}
\put(198,266){\makebox(0,0)[r]{20}}
\put(1416.0,266.0){\rule[-0.200pt]{4.818pt}{0.400pt}}
\put(220.0,419.0){\rule[-0.200pt]{4.818pt}{0.400pt}}
\put(198,419){\makebox(0,0)[r]{40}}
\put(1416.0,419.0){\rule[-0.200pt]{4.818pt}{0.400pt}}
\put(220.0,571.0){\rule[-0.200pt]{4.818pt}{0.400pt}}
\put(198,571){\makebox(0,0)[r]{60}}
\put(1416.0,571.0){\rule[-0.200pt]{4.818pt}{0.400pt}}
\put(220.0,724.0){\rule[-0.200pt]{4.818pt}{0.400pt}}
\put(198,724){\makebox(0,0)[r]{80}}
\put(1416.0,724.0){\rule[-0.200pt]{4.818pt}{0.400pt}}
\put(220.0,877.0){\rule[-0.200pt]{4.818pt}{0.400pt}}
\put(198,877){\makebox(0,0)[r]{100}}
\put(1416.0,877.0){\rule[-0.200pt]{4.818pt}{0.400pt}}
\put(388.0,113.0){\rule[-0.200pt]{0.400pt}{4.818pt}}
\put(388,68){\makebox(0,0){50}}
\put(388.0,857.0){\rule[-0.200pt]{0.400pt}{4.818pt}}
\put(597.0,113.0){\rule[-0.200pt]{0.400pt}{4.818pt}}
\put(597,68){\makebox(0,0){100}}
\put(597.0,857.0){\rule[-0.200pt]{0.400pt}{4.818pt}}
\put(807.0,113.0){\rule[-0.200pt]{0.400pt}{4.818pt}}
\put(807,68){\makebox(0,0){150}}
\put(807.0,857.0){\rule[-0.200pt]{0.400pt}{4.818pt}}
\put(1017.0,113.0){\rule[-0.200pt]{0.400pt}{4.818pt}}
\put(1017,68){\makebox(0,0){200}}
\put(1017.0,857.0){\rule[-0.200pt]{0.400pt}{4.818pt}}
\put(1226.0,113.0){\rule[-0.200pt]{0.400pt}{4.818pt}}
\put(1226,68){\makebox(0,0){250}}
\put(1226.0,857.0){\rule[-0.200pt]{0.400pt}{4.818pt}}
\put(1436.0,113.0){\rule[-0.200pt]{0.400pt}{4.818pt}}
\put(1436,68){\makebox(0,0){300}}
\put(1436.0,857.0){\rule[-0.200pt]{0.400pt}{4.818pt}}
\put(220.0,113.0){\rule[-0.200pt]{292.934pt}{0.400pt}}
\put(1436.0,113.0){\rule[-0.200pt]{0.400pt}{184.048pt}}
\put(220.0,877.0){\rule[-0.200pt]{292.934pt}{0.400pt}}
\put(20,495){\makebox(0,0) {$\delta a_{\mu}^{tr} \times 10^{-10}$}}
\put(828,-20){\makebox(0,0){$M_Y$(GeV)}}
\put(220.0,113.0){\rule[-0.200pt]{0.400pt}{184.048pt}}
\put(1306,812){\makebox(0,0)[r] {$\delta a_{\mu}^{tr}$}}
\multiput(398.59,865.24)(0.482,-3.655){9}{\rule{0.116pt}{2.833pt}}
\multiput(397.17,871.12)(6.000,-35.119){2}{\rule{0.400pt}{1.417pt}}
\multiput(404.58,827.03)(0.493,-2.638){23}{\rule{0.119pt}{2.162pt}}
\multiput(403.17,831.51)(13.000,-62.514){2}{\rule{0.400pt}{1.081pt}}
\multiput(417.58,760.84)(0.492,-2.392){21}{\rule{0.119pt}{1.967pt}}
\multiput(416.17,764.92)(12.000,-51.918){2}{\rule{0.400pt}{0.983pt}}
\multiput(429.58,705.80)(0.492,-2.090){21}{\rule{0.119pt}{1.733pt}}
\multiput(428.17,709.40)(12.000,-45.402){2}{\rule{0.400pt}{0.867pt}}
\multiput(441.58,657.63)(0.492,-1.832){21}{\rule{0.119pt}{1.533pt}}
\multiput(440.17,660.82)(12.000,-39.817){2}{\rule{0.400pt}{0.767pt}}
\multiput(453.58,615.86)(0.493,-1.448){23}{\rule{0.119pt}{1.238pt}}
\multiput(452.17,618.43)(13.000,-34.430){2}{\rule{0.400pt}{0.619pt}}
\multiput(466.58,579.02)(0.492,-1.401){21}{\rule{0.119pt}{1.200pt}}
\multiput(465.17,581.51)(12.000,-30.509){2}{\rule{0.400pt}{0.600pt}}
\multiput(478.58,546.57)(0.492,-1.229){21}{\rule{0.119pt}{1.067pt}}
\multiput(477.17,548.79)(12.000,-26.786){2}{\rule{0.400pt}{0.533pt}}
\multiput(490.58,518.26)(0.493,-1.012){23}{\rule{0.119pt}{0.900pt}}
\multiput(489.17,520.13)(13.000,-24.132){2}{\rule{0.400pt}{0.450pt}}
\multiput(503.58,492.40)(0.492,-0.970){21}{\rule{0.119pt}{0.867pt}}
\multiput(502.17,494.20)(12.000,-21.201){2}{\rule{0.400pt}{0.433pt}}
\multiput(515.58,469.82)(0.492,-0.841){21}{\rule{0.119pt}{0.767pt}}
\multiput(514.17,471.41)(12.000,-18.409){2}{\rule{0.400pt}{0.383pt}}
\multiput(527.58,449.96)(0.492,-0.798){21}{\rule{0.119pt}{0.733pt}}
\multiput(526.17,451.48)(12.000,-17.478){2}{\rule{0.400pt}{0.367pt}}
\multiput(539.58,431.41)(0.493,-0.655){23}{\rule{0.119pt}{0.623pt}}
\multiput(538.17,432.71)(13.000,-15.707){2}{\rule{0.400pt}{0.312pt}}
\multiput(552.58,414.51)(0.492,-0.625){21}{\rule{0.119pt}{0.600pt}}
\multiput(551.17,415.75)(12.000,-13.755){2}{\rule{0.400pt}{0.300pt}}
\multiput(564.58,399.65)(0.492,-0.582){21}{\rule{0.119pt}{0.567pt}}
\multiput(563.17,400.82)(12.000,-12.824){2}{\rule{0.400pt}{0.283pt}}
\multiput(576.00,386.92)(0.496,-0.492){21}{\rule{0.500pt}{0.119pt}}
\multiput(576.00,387.17)(10.962,-12.000){2}{\rule{0.250pt}{0.400pt}}
\multiput(588.00,374.92)(0.539,-0.492){21}{\rule{0.533pt}{0.119pt}}
\multiput(588.00,375.17)(11.893,-12.000){2}{\rule{0.267pt}{0.400pt}}
\multiput(601.00,362.92)(0.600,-0.491){17}{\rule{0.580pt}{0.118pt}}
\multiput(601.00,363.17)(10.796,-10.000){2}{\rule{0.290pt}{0.400pt}}
\multiput(613.00,352.92)(0.600,-0.491){17}{\rule{0.580pt}{0.118pt}}
\multiput(613.00,353.17)(10.796,-10.000){2}{\rule{0.290pt}{0.400pt}}
\multiput(625.00,342.93)(0.728,-0.489){15}{\rule{0.678pt}{0.118pt}}
\multiput(625.00,343.17)(11.593,-9.000){2}{\rule{0.339pt}{0.400pt}}
\multiput(638.00,333.93)(0.758,-0.488){13}{\rule{0.700pt}{0.117pt}}
\multiput(638.00,334.17)(10.547,-8.000){2}{\rule{0.350pt}{0.400pt}}
\multiput(650.00,325.93)(0.758,-0.488){13}{\rule{0.700pt}{0.117pt}}
\multiput(650.00,326.17)(10.547,-8.000){2}{\rule{0.350pt}{0.400pt}}
\multiput(662.00,317.93)(0.874,-0.485){11}{\rule{0.786pt}{0.117pt}}
\multiput(662.00,318.17)(10.369,-7.000){2}{\rule{0.393pt}{0.400pt}}
\multiput(674.00,310.93)(1.123,-0.482){9}{\rule{0.967pt}{0.116pt}}
\multiput(674.00,311.17)(10.994,-6.000){2}{\rule{0.483pt}{0.400pt}}
\multiput(687.00,304.93)(1.033,-0.482){9}{\rule{0.900pt}{0.116pt}}
\multiput(687.00,305.17)(10.132,-6.000){2}{\rule{0.450pt}{0.400pt}}
\multiput(699.00,298.93)(1.033,-0.482){9}{\rule{0.900pt}{0.116pt}}
\multiput(699.00,299.17)(10.132,-6.000){2}{\rule{0.450pt}{0.400pt}}
\multiput(711.00,292.93)(1.378,-0.477){7}{\rule{1.140pt}{0.115pt}}
\multiput(711.00,293.17)(10.634,-5.000){2}{\rule{0.570pt}{0.400pt}}
\multiput(724.00,287.93)(1.267,-0.477){7}{\rule{1.060pt}{0.115pt}}
\multiput(724.00,288.17)(9.800,-5.000){2}{\rule{0.530pt}{0.400pt}}
\multiput(736.00,282.93)(1.267,-0.477){7}{\rule{1.060pt}{0.115pt}}
\multiput(736.00,283.17)(9.800,-5.000){2}{\rule{0.530pt}{0.400pt}}
\multiput(748.00,277.94)(1.651,-0.468){5}{\rule{1.300pt}{0.113pt}}
\multiput(748.00,278.17)(9.302,-4.000){2}{\rule{0.650pt}{0.400pt}}
\multiput(760.00,273.94)(1.797,-0.468){5}{\rule{1.400pt}{0.113pt}}
\multiput(760.00,274.17)(10.094,-4.000){2}{\rule{0.700pt}{0.400pt}}
\multiput(773.00,269.94)(1.651,-0.468){5}{\rule{1.300pt}{0.113pt}}
\multiput(773.00,270.17)(9.302,-4.000){2}{\rule{0.650pt}{0.400pt}}
\multiput(785.00,265.94)(1.651,-0.468){5}{\rule{1.300pt}{0.113pt}}
\multiput(785.00,266.17)(9.302,-4.000){2}{\rule{0.650pt}{0.400pt}}
\multiput(797.00,261.95)(2.695,-0.447){3}{\rule{1.833pt}{0.108pt}}
\multiput(797.00,262.17)(9.195,-3.000){2}{\rule{0.917pt}{0.400pt}}
\multiput(810.00,258.95)(2.472,-0.447){3}{\rule{1.700pt}{0.108pt}}
\multiput(810.00,259.17)(8.472,-3.000){2}{\rule{0.850pt}{0.400pt}}
\multiput(822.00,255.95)(2.472,-0.447){3}{\rule{1.700pt}{0.108pt}}
\multiput(822.00,256.17)(8.472,-3.000){2}{\rule{0.850pt}{0.400pt}}
\multiput(834.00,252.95)(2.472,-0.447){3}{\rule{1.700pt}{0.108pt}}
\multiput(834.00,253.17)(8.472,-3.000){2}{\rule{0.850pt}{0.400pt}}
\multiput(846.00,249.95)(2.695,-0.447){3}{\rule{1.833pt}{0.108pt}}
\multiput(846.00,250.17)(9.195,-3.000){2}{\rule{0.917pt}{0.400pt}}
\put(859,246.17){\rule{2.500pt}{0.400pt}}
\multiput(859.00,247.17)(6.811,-2.000){2}{\rule{1.250pt}{0.400pt}}
\multiput(871.00,244.95)(2.472,-0.447){3}{\rule{1.700pt}{0.108pt}}
\multiput(871.00,245.17)(8.472,-3.000){2}{\rule{0.850pt}{0.400pt}}
\put(883,241.17){\rule{2.700pt}{0.400pt}}
\multiput(883.00,242.17)(7.396,-2.000){2}{\rule{1.350pt}{0.400pt}}
\put(896,239.17){\rule{2.500pt}{0.400pt}}
\multiput(896.00,240.17)(6.811,-2.000){2}{\rule{1.250pt}{0.400pt}}
\put(908,237.17){\rule{2.500pt}{0.400pt}}
\multiput(908.00,238.17)(6.811,-2.000){2}{\rule{1.250pt}{0.400pt}}
\put(920,235.17){\rule{2.500pt}{0.400pt}}
\multiput(920.00,236.17)(6.811,-2.000){2}{\rule{1.250pt}{0.400pt}}
\put(932,233.17){\rule{2.700pt}{0.400pt}}
\multiput(932.00,234.17)(7.396,-2.000){2}{\rule{1.350pt}{0.400pt}}
\put(945,231.17){\rule{2.500pt}{0.400pt}}
\multiput(945.00,232.17)(6.811,-2.000){2}{\rule{1.250pt}{0.400pt}}
\put(957,229.17){\rule{2.500pt}{0.400pt}}
\multiput(957.00,230.17)(6.811,-2.000){2}{\rule{1.250pt}{0.400pt}}
\put(969,227.67){\rule{3.132pt}{0.400pt}}
\multiput(969.00,228.17)(6.500,-1.000){2}{\rule{1.566pt}{0.400pt}}
\put(982,226.17){\rule{2.500pt}{0.400pt}}
\multiput(982.00,227.17)(6.811,-2.000){2}{\rule{1.250pt}{0.400pt}}
\put(994,224.67){\rule{2.891pt}{0.400pt}}
\multiput(994.00,225.17)(6.000,-1.000){2}{\rule{1.445pt}{0.400pt}}
\put(1006,223.17){\rule{2.500pt}{0.400pt}}
\multiput(1006.00,224.17)(6.811,-2.000){2}{\rule{1.250pt}{0.400pt}}
\put(1018,221.67){\rule{3.132pt}{0.400pt}}
\multiput(1018.00,222.17)(6.500,-1.000){2}{\rule{1.566pt}{0.400pt}}
\put(1031,220.17){\rule{2.500pt}{0.400pt}}
\multiput(1031.00,221.17)(6.811,-2.000){2}{\rule{1.250pt}{0.400pt}}
\put(1043,218.67){\rule{2.891pt}{0.400pt}}
\multiput(1043.00,219.17)(6.000,-1.000){2}{\rule{1.445pt}{0.400pt}}
\put(1055,217.67){\rule{3.132pt}{0.400pt}}
\multiput(1055.00,218.17)(6.500,-1.000){2}{\rule{1.566pt}{0.400pt}}
\put(1068,216.67){\rule{2.891pt}{0.400pt}}
\multiput(1068.00,217.17)(6.000,-1.000){2}{\rule{1.445pt}{0.400pt}}
\put(1080,215.67){\rule{2.891pt}{0.400pt}}
\multiput(1080.00,216.17)(6.000,-1.000){2}{\rule{1.445pt}{0.400pt}}
\put(1092,214.67){\rule{2.891pt}{0.400pt}}
\multiput(1092.00,215.17)(6.000,-1.000){2}{\rule{1.445pt}{0.400pt}}
\put(1104,213.67){\rule{3.132pt}{0.400pt}}
\multiput(1104.00,214.17)(6.500,-1.000){2}{\rule{1.566pt}{0.400pt}}
\put(1117,212.67){\rule{2.891pt}{0.400pt}}
\multiput(1117.00,213.17)(6.000,-1.000){2}{\rule{1.445pt}{0.400pt}}
\put(1129,211.67){\rule{2.891pt}{0.400pt}}
\multiput(1129.00,212.17)(6.000,-1.000){2}{\rule{1.445pt}{0.400pt}}
\put(1141,210.67){\rule{2.891pt}{0.400pt}}
\multiput(1141.00,211.17)(6.000,-1.000){2}{\rule{1.445pt}{0.400pt}}
\put(1153,209.67){\rule{3.132pt}{0.400pt}}
\multiput(1153.00,210.17)(6.500,-1.000){2}{\rule{1.566pt}{0.400pt}}
\put(1166,208.67){\rule{2.891pt}{0.400pt}}
\multiput(1166.00,209.17)(6.000,-1.000){2}{\rule{1.445pt}{0.400pt}}
\put(1178,207.67){\rule{2.891pt}{0.400pt}}
\multiput(1178.00,208.17)(6.000,-1.000){2}{\rule{1.445pt}{0.400pt}}
\put(1190,206.67){\rule{3.132pt}{0.400pt}}
\multiput(1190.00,207.17)(6.500,-1.000){2}{\rule{1.566pt}{0.400pt}}
\put(1328.0,812.0){\rule[-0.200pt]{15.899pt}{0.400pt}}
\put(1215,205.67){\rule{2.891pt}{0.400pt}}
\multiput(1215.00,206.17)(6.000,-1.000){2}{\rule{1.445pt}{0.400pt}}
\put(1227,204.67){\rule{2.891pt}{0.400pt}}
\multiput(1227.00,205.17)(6.000,-1.000){2}{\rule{1.445pt}{0.400pt}}
\put(1203.0,207.0){\rule[-0.200pt]{2.891pt}{0.400pt}}
\put(1252,203.67){\rule{2.891pt}{0.400pt}}
\multiput(1252.00,204.17)(6.000,-1.000){2}{\rule{1.445pt}{0.400pt}}
\put(1264,202.67){\rule{2.891pt}{0.400pt}}
\multiput(1264.00,203.17)(6.000,-1.000){2}{\rule{1.445pt}{0.400pt}}
\put(1239.0,205.0){\rule[-0.200pt]{3.132pt}{0.400pt}}
\put(1289,201.67){\rule{2.891pt}{0.400pt}}
\multiput(1289.00,202.17)(6.000,-1.000){2}{\rule{1.445pt}{0.400pt}}
\put(1301,200.67){\rule{2.891pt}{0.400pt}}
\multiput(1301.00,201.17)(6.000,-1.000){2}{\rule{1.445pt}{0.400pt}}
\put(1276.0,203.0){\rule[-0.200pt]{3.132pt}{0.400pt}}
\put(1325,199.67){\rule{3.132pt}{0.400pt}}
\multiput(1325.00,200.17)(6.500,-1.000){2}{\rule{1.566pt}{0.400pt}}
\put(1313.0,201.0){\rule[-0.200pt]{2.891pt}{0.400pt}}
\put(1350,198.67){\rule{2.891pt}{0.400pt}}
\multiput(1350.00,199.17)(6.000,-1.000){2}{\rule{1.445pt}{0.400pt}}
\put(1338.0,200.0){\rule[-0.200pt]{2.891pt}{0.400pt}}
\put(1375,197.67){\rule{2.891pt}{0.400pt}}
\multiput(1375.00,198.17)(6.000,-1.000){2}{\rule{1.445pt}{0.400pt}}
\put(1362.0,199.0){\rule[-0.200pt]{3.132pt}{0.400pt}}
\put(1399,196.67){\rule{2.891pt}{0.400pt}}
\multiput(1399.00,197.17)(6.000,-1.000){2}{\rule{1.445pt}{0.400pt}}
\put(1387.0,198.0){\rule[-0.200pt]{2.891pt}{0.400pt}}
\put(1411.0,197.0){\rule[-0.200pt]{6.022pt}{0.400pt}}
\put(1306,767){\makebox(0,0)[r]{Uper limit}}

\multiput(1328,767)(20.756,0.000){4}{\usebox{\plotpoint}}
\put(1394,767){\usebox{\plotpoint}}
\put(220,564){\usebox{\plotpoint}}
\put(220.00,564.00){\usebox{\plotpoint}}
\put(240.76,564.00){\usebox{\plotpoint}}
\multiput(245,564)(20.756,0.000){0}{\usebox{\plotpoint}}
\put(261.51,564.00){\usebox{\plotpoint}}
\multiput(269,564)(20.756,0.000){0}{\usebox{\plotpoint}}
\put(282.27,564.00){\usebox{\plotpoint}}
\put(303.02,564.00){\usebox{\plotpoint}}
\multiput(306,564)(20.756,0.000){0}{\usebox{\plotpoint}}
\put(323.78,564.00){\usebox{\plotpoint}}
\multiput(331,564)(20.756,0.000){0}{\usebox{\plotpoint}}
\put(344.53,564.00){\usebox{\plotpoint}}
\put(365.29,564.00){\usebox{\plotpoint}}
\multiput(367,564)(20.756,0.000){0}{\usebox{\plotpoint}}
\put(386.04,564.00){\usebox{\plotpoint}}
\multiput(392,564)(20.756,0.000){0}{\usebox{\plotpoint}}
\put(406.80,564.00){\usebox{\plotpoint}}
\put(427.55,564.00){\usebox{\plotpoint}}
\multiput(429,564)(20.756,0.000){0}{\usebox{\plotpoint}}
\put(448.31,564.00){\usebox{\plotpoint}}
\multiput(453,564)(20.756,0.000){0}{\usebox{\plotpoint}}
\put(469.07,564.00){\usebox{\plotpoint}}
\put(489.82,564.00){\usebox{\plotpoint}}
\multiput(490,564)(20.756,0.000){0}{\usebox{\plotpoint}}
\put(510.58,564.00){\usebox{\plotpoint}}
\multiput(515,564)(20.756,0.000){0}{\usebox{\plotpoint}}
\put(531.33,564.00){\usebox{\plotpoint}}
\multiput(539,564)(20.756,0.000){0}{\usebox{\plotpoint}}
\put(552.09,564.00){\usebox{\plotpoint}}
\put(572.84,564.00){\usebox{\plotpoint}}
\multiput(576,564)(20.756,0.000){0}{\usebox{\plotpoint}}
\put(593.60,564.00){\usebox{\plotpoint}}
\multiput(601,564)(20.756,0.000){0}{\usebox{\plotpoint}}
\put(614.35,564.00){\usebox{\plotpoint}}
\put(635.11,564.00){\usebox{\plotpoint}}
\multiput(638,564)(20.756,0.000){0}{\usebox{\plotpoint}}
\put(655.87,564.00){\usebox{\plotpoint}}
\multiput(662,564)(20.756,0.000){0}{\usebox{\plotpoint}}
\put(676.62,564.00){\usebox{\plotpoint}}
\put(697.38,564.00){\usebox{\plotpoint}}
\multiput(699,564)(20.756,0.000){0}{\usebox{\plotpoint}}
\put(718.13,564.00){\usebox{\plotpoint}}
\multiput(724,564)(20.756,0.000){0}{\usebox{\plotpoint}}
\put(738.89,564.00){\usebox{\plotpoint}}
\put(759.64,564.00){\usebox{\plotpoint}}
\multiput(760,564)(20.756,0.000){0}{\usebox{\plotpoint}}
\put(780.40,564.00){\usebox{\plotpoint}}
\multiput(785,564)(20.756,0.000){0}{\usebox{\plotpoint}}
\put(801.15,564.00){\usebox{\plotpoint}}
\put(821.91,564.00){\usebox{\plotpoint}}
\multiput(822,564)(20.756,0.000){0}{\usebox{\plotpoint}}
\put(842.66,564.00){\usebox{\plotpoint}}
\multiput(846,564)(20.756,0.000){0}{\usebox{\plotpoint}}
\put(863.42,564.00){\usebox{\plotpoint}}
\multiput(871,564)(20.756,0.000){0}{\usebox{\plotpoint}}
\put(884.18,564.00){\usebox{\plotpoint}}
\put(904.93,564.00){\usebox{\plotpoint}}
\multiput(908,564)(20.756,0.000){0}{\usebox{\plotpoint}}
\put(925.69,564.00){\usebox{\plotpoint}}
\multiput(932,564)(20.756,0.000){0}{\usebox{\plotpoint}}
\put(946.44,564.00){\usebox{\plotpoint}}
\put(967.20,564.00){\usebox{\plotpoint}}
\multiput(969,564)(20.756,0.000){0}{\usebox{\plotpoint}}
\put(987.95,564.00){\usebox{\plotpoint}}
\multiput(994,564)(20.756,0.000){0}{\usebox{\plotpoint}}
\put(1008.71,564.00){\usebox{\plotpoint}}
\put(1029.46,564.00){\usebox{\plotpoint}}
\multiput(1031,564)(20.756,0.000){0}{\usebox{\plotpoint}}
\put(1050.22,564.00){\usebox{\plotpoint}}
\multiput(1055,564)(20.756,0.000){0}{\usebox{\plotpoint}}
\put(1070.98,564.00){\usebox{\plotpoint}}
\put(1091.73,564.00){\usebox{\plotpoint}}
\multiput(1092,564)(20.756,0.000){0}{\usebox{\plotpoint}}
\put(1112.49,564.00){\usebox{\plotpoint}}
\multiput(1117,564)(20.756,0.000){0}{\usebox{\plotpoint}}
\put(1133.24,564.00){\usebox{\plotpoint}}
\multiput(1141,564)(20.756,0.000){0}{\usebox{\plotpoint}}
\put(1154.00,564.00){\usebox{\plotpoint}}
\put(1174.75,564.00){\usebox{\plotpoint}}
\multiput(1178,564)(20.756,0.000){0}{\usebox{\plotpoint}}
\put(1195.51,564.00){\usebox{\plotpoint}}
\multiput(1203,564)(20.756,0.000){0}{\usebox{\plotpoint}}
\put(1216.26,564.00){\usebox{\plotpoint}}
\put(1237.02,564.00){\usebox{\plotpoint}}
\multiput(1239,564)(20.756,0.000){0}{\usebox{\plotpoint}}
\put(1257.77,564.00){\usebox{\plotpoint}}
\multiput(1264,564)(20.756,0.000){0}{\usebox{\plotpoint}}
\put(1278.53,564.00){\usebox{\plotpoint}}
\put(1299.29,564.00){\usebox{\plotpoint}}
\multiput(1301,564)(20.756,0.000){0}{\usebox{\plotpoint}}
\put(1320.04,564.00){\usebox{\plotpoint}}
\multiput(1325,564)(20.756,0.000){0}{\usebox{\plotpoint}}
\put(1340.80,564.00){\usebox{\plotpoint}}
\put(1361.55,564.00){\usebox{\plotpoint}}
\multiput(1362,564)(20.756,0.000){0}{\usebox{\plotpoint}}
\put(1382.31,564.00){\usebox{\plotpoint}}
\multiput(1387,564)(20.756,0.000){0}{\usebox{\plotpoint}}
\put(1403.06,564.00){\usebox{\plotpoint}}
\put(1423.82,564.00){\usebox{\plotpoint}}
\multiput(1424,564)(20.756,0.000){0}{\usebox{\plotpoint}}
\put(1436,564){\usebox{\plotpoint}}
\sbox{\plotpoint}{\rule[-0.400pt]{0.800pt}{0.800pt}}%
\put(1306,722){\makebox(0,0)[r]{Lower limit}}


\put(1328.0,722.0){\rule[-0.400pt]{15.899pt}{0.800pt}}
\put(220,319){\usebox{\plotpoint}}
\put(220.0,319.0){\rule[-0.400pt]{292.934pt}{0.800pt}}
\end{picture}